
\input mssymb

\def\chapter#1#2{\vskip0pt plus .3\vsize\penalty-250
\vskip0pt plus -.3\vsize\bigskip\vskip\parskip
\message {#1 #2}
\centerline{\titling #1}
\centerline{\titling #2}
\nobreak\bigskip\noindent}

\magnification=1200
\font\titling=cmr17
\font\proclaiming=cmr12
\font\ninesl=cmsl9

\def\Z{{\Bbb Z}}

\def\B{{\cal B}}

\def\Cyl #1#2{[#1]^{#2}}
\def\Dcyl #1#2#3{[#1]_{#2}^{#3}}

\def\E{{\cal E}}

\def\R{{\Bbb R}}
\def\N{{\Bbb N}}

\def\PP{{\Bbb P}}
\def\G{{\cal G}}

\outer\def\proclaim#1. #2\endproclaim
{\medbreak\noindent{\proclaiming#1.}
\enspace{\sl#2}\par\ifdim\lastskip<\medskipamount
\removelastskip\penalty55\medskip\fi}

\outer\def\proc#1
{\medbreak\noindent{\bf#1.}
\enspace\sl}

\def\endproc
{\rm\par\ifdim\lastskip<\medskipamount\removelastskip\penalty55\medskip\fi}

\outer\def\demo#1{\medbreak\noindent{\bf#1.}}
\def\enddemo{\medbreak}
\def\ra{\rightarrow}

\def\Ra{\Rightarrow}
\def\Lra{\Leftrightarrow}

\def\sf#1#2{{{#1}\over{#2}}}

\def\Abstract#1{\centerline{{\bf Abstract}}\medskip
\hbox to \hsize{\hss\vbox{\hsize 23pc\noindent#1}\hss}}

\font\ninerm=cmr9

\def\demo#1{\medbreak\noindent{\sl #1.}}

\newcount\chno
\newcount\thno
\newcount\defno
\newcount\tchno
\newcount\tno
\newcount\eqnum
\newcount\secno
\newcount\rnum
\global\rnum=0
\global\chno=1
\global\thno=1
\global\defno=1
\global\tchno=1
\global\eqnum=0
\global\secno=1

\def \eqgdef{\tchno =1\tno =1(\ref)}
\def \eqgcons{\tchno =1\tno =2(\ref)}

\def \gtomap{\tchno =1\tno =2\ref}
\def \eqgclass{\tchno =1\tno =3(\ref)}
\def \eqintg{\tchno =1\tno =4(\ref)}
\def \eqTderiv{\tchno =1\tno =5(\ref)}
\def \hmeas{\tchno =1\tno =3\ref}
\def \eqhclass{\tchno =1\tno =6(\ref)}
\def \eqevo{\tchno =1\tno =7(\ref)}
\def \eqhdef{\tchno =1\tno =8(\ref)}

\def\ref{\relax\ifnum \tchno=\chno
\the\tno
\else
\the\tchno$\cdot$\the\tno
\fi
}

\def\rref{\the\tno}

\def\Defn
{\medbreak\noindent{\bf Definition \number\defno.}
\global\advance\defno by 1
\enspace\sl}

\def\Theorem
{\medbreak\noindent{\bf Theorem \number\thno.}
\global\advance\thno by 1
\enspace\sl}

\def\Lemma
{\medbreak\noindent{\bf Lemma \number\thno.}
\global\advance\thno by 1
\enspace\sl}

\def\Prop
{\medbreak\noindent{\bf Proposition \number\thno.}
\global\advance\thno by 1
\enspace\sl}

\outer\def\Section #1\par
{
\vskip0pt plus .3\vsize\penalty -250
\vskip0pt plus -.3\vsize\bigskip\vskip\parskip
\message{Section \number\chno.\number\secno }
\leftline{\bf \number\secno. #1}
\nobreak\smallskip\noindent
\global\advance\secno by 1
}

\def\newch
{
\global\advance\chno by 1
\global\thno=1
\global\defno=1
\global\eqnum=0
\global\secno=1
}

\def\endproc
{\rm\par\ifdim\lastskip<\medskipamount\removelastskip\penalty55\medskip\fi}


\hbox to \hsize{\hss
{\bf NON-ERGODICITY FOR C\raise0.8ex\hbox{\ninerm 1} EXPANDING MAPS}\hss}
\bigskip
\centerline{Anthony N. Quas}
\medskip
\centerline{{\ninesl Mathematics Institute, University of Warwick,
Coventry, CV4 7AL, U.K.}}
\bigskip
\bigskip

In this paper, we use a remarkable recent result of Bramson and Kalikow, [BK],
to produce an example of a $C^1$ expanding map of the circle which preserves
more than one absolutely continuous probability measure.
\footnote{}{{\ninesl 1991 Mathematics Subject Classification.}
{\ninerm Primary 58F11; Secondary 28D05.}}
\footnote{}{{\ninesl Key words and phrases.}
{\ninerm Expanding maps, absolutely continuous invariant measures}}
\footnote{}{{\ninerm Research supported by a SERC grant}}
We first introduce some abbreviations and definitions.

An {\it expanding} map of the circle is a differentiable map $f$ from
the circle to itself such that $|D_x f^n|\geq C$ for some fixed constants
$C>1$,
$n\in\N$, for all $x\in S^1$. Such
a map must be an $r$-fold cover of the circle for some $r\geq 2$.
We will denote by ${\cal E}^k$, the
collection of $C^k$ expanding maps of the circle, where
$k\geq 1$. Note that if $k$ is non-integral, then we mean that the
$\lfloor k\rfloor$th derivative of the map is H\"older
with exponent $k-\lfloor k\rfloor$.

An {\it absolutely continuous invariant probability measure} (abbreviated to
ACIM) for a map
$T:S^1\ra S^1$ is a Borel probability measure $\mu$ such that
$\mu(T^{-1}B)=\mu(B)$ for all Borel sets $B$ and such that $\mu$ is
absolutely continuous with respect to Lebesgue measure $\lambda$
on $S^1$ (that is $\lambda(A)=0\Ra\mu(A)=0$).

The example which we describe will make use of symbolic dynamics. The
central object of the investigation will be the set
$\Sigma_s=\{0,1,\ldots s-1\}^{\Z^+}$.
This is endowed with the product topology by taking
the metric
$$
d(x,y)=\cases {
0 &if $x=y$\cr
2^{-n} &if $x_i=y_i$ for $i=0,\ldots,n-1$, but $x_n\neq y_n$.
}
$$
In fact, we will only consider the space $\Sigma_2$ in the remainder of this
paper.
We will then consider those Borel probability measures on $\Sigma_2$
which are invariant
under the shift map $\sigma:\Sigma_2\ra\Sigma_2,\ \sigma(x)_n=x_{n+1}$. This
map may be easily seen to be continuous. Such measures are called
{\it shift-invariant}. We will consider
a particular class of such measures known as $g$-measures.
If $x\in\Sigma_2$, and $a=(a_0,a_1,\ldots,a_{s-1})\in\{0,1\}^s$
is a finite word (possibly
empty), then by $ax$, we mean the sequence in $\Sigma_2$ with
first $s$ terms, those of $a$ and succeeding terms those of $x$, that is
$$
(ax)_i=\cases{
a_i&if $i<s$;\cr
x_{i-s}&if $i\geq s$.
}$$

If $x\in \Sigma_2$, then we define $\Cyl xn$ to be the
{\it $n$th cylinder} about $x$: $\Cyl xn=\{y:d(x,y)<2^{-n}\}$.
Let $g$ be a continuous function $\Sigma_2\ra(0,1)$ such that
$g(0x)+g(1x)=1$ for all $x\in\Sigma_2$. Write $\G$ for the set of all such
functions. A measure $m$ is called a
{\it $g$-measure} if it is shift-invariant and it satisfies
$$
m\bigl(\Cyl x0|\sigma^{-1}\B\bigr)(x)=g(x){\rm\ for\ almost\ all\ }x\in
\Sigma_2{\rm\ (wrt.\ }m),
$$
where $\B$ is the Borel $\sigma$-algebra of the space $\Sigma_2$.
This is the statement that the conditional probability of the first term
of a sequence being $x_0$ given that the succeeding terms are
$x_1,x_2,\ldots$ is given by $g(x)$. By applying some theorems of martingale
theory, we can deduce the following equivalent statement:
$m$ is a $g$-measure if and only if
$m(\Cyl xn)>0$ for each $x\in\Sigma_2$, $n\in\Z^+$, and
$$
\lim_{n\ra\infty} {{m(\Cyl xn)}\over{m(\Cyl {\sigma(x)}{n-1})}}=g(x)
{\rm \ for\ almost\ all\ }x\in\Sigma_2\ ({\rm wrt.\ }m).
$$
Note that each $g\in\G$ must be strictly bounded away from 0 and 1, since
the space $\Sigma_2$ is compact.
For every $g\in\G$, there is always at least one $g$-measure by a
standard ergodic theory argument.
A $g$-measure is always non-atomic and of full support.
For proofs of these facts and other background information,
the reader is referred to [Ke]. Note that the $g$-measures form a convex
set. The extreme points of this set are ergodic. It follows that for a
particular $g$, there is more than one $g$-measure if and only if there
exists a non-ergodic $g$-measure.

Let $f$ be a continuous function $\Sigma_2\ra\R$.
The {\it $k$th variation} of $f$
is ${\rm var}_k(f)=\max\{|f(x)-f(y)|:x,y\in \Sigma_2;\ d(x,y)<2^{-k}\}$.
Note ${\rm var}_k(f)\ra 0$ as $k\ra\infty$.
The function $f$ is defined to be {\it H\"older
continuous} if ${\rm var}_k(f)<C\theta^k$ for some $C>0$ and $0<\theta<1$,
for all $k\geq 0$.

Keane ([Ke]) showed that if $g\in\G$ satisfies strong continuity
properties, then there is a unique $g$-measure. Walters ([W]) improved
this, showing that if $g$ has summable variation, or in particular, if $g$
is H\"older continuous, then there
is a unique $g$-measure. They asked whether this result could be extended to
the case of $g$ continuous, without any further restriction. This
question has attracted a very large amount of attention over the past 20 years
with some recent partial results obtained in [H] and [B].
It has recently been shown by Bramson and Kalikow (see [BK]) that there
exist examples of continuous $g\in\G$ for which there exist more than one
$g$-measure, thus solving the problem.

This development largely mirrors the development of results concerning the
existence of ACIMs for expanding maps of the circle. It was shown by
Krzy\.zewski and Szlenk (see [KS] and [Kr1])
that if $f$ is in $\E^k$ with $k\geq 2$,
then $f$ has a unique ACIM. A steady stream of results has continued since
the publication of this paper resulting in a refined version of
the result being known as the
`Folklore Theorem'. In [Kr2], it was shown that there exists $f\in\E^1$ for
which there is no ACIM, and an explicit example was constructed in [GS].
This however still leaves the question: Does there
exist $f\in\E^1$ for which there is more than one ACIM? It is this question
which we answer in this paper, showing that there is an $f\in\E^1$ which
preserves Lebesgue measure, but for which Lebesgue measure is not ergodic.

\Theorem
There exists a $C^1$ expanding map of the circle which preserves Lebesgue
measure $\lambda$, but for which $\lambda$ is not ergodic.
\endproc

The proof of the theorem is in three steps, which are separated out as
lemmas. The third step is an almost
exact mimicking of the proof given in [BK].

We start by defining an equivalence relation $\sim$ on $\Sigma_2$.
$$
x\sim y\rm{\ if\ }\cases
{x=y,\cr
x=a0111\ldots{\rm\ and\ }y=a1000\ldots&where $a$ is a finite sequence,\cr
x=a1000\ldots{\rm\ and\ }y=a0111\ldots&where $a$ is a finite sequence,\cr
x=0000\ldots{\rm\ and\ }y=1111\ldots&or\cr
x=1111\ldots{\rm\ and\ }y=0000\ldots}
$$
Note that by a finite sequence, we are also allowing the possibility that
the sequence be empty.

\Lemma
Suppose $g\in\G$ and $m$ is a $g$-measure. Suppose further that $g$ has the
property that
$$x\sim y \Ra g(x)=g(y).
$$
Then there is a continuous surjection $\pi:\Sigma_2\ra S^1$ and an expanding
$C^1$ map $T:S^1\ra S^1$ such that $\pi$ is a measure-theoretic isomorphism
between $(\sigma,m)$ and $(T,\lambda)$.
\endproc

\demo{Proof}
Define a total order
$\preceq$ on $\Sigma_2$, the lexicographic ordering:
$$
x\prec y\Lra\exists n\geq 0{\rm\ such\ that\ }x_0=y_0,\ldots,x_{n-1}=y_{n-1}
{\rm\ and\ }x_n<y_n.
$$
Now, set $[x,y]=\{z:x\preceq z\preceq y\}$ and define the open intervals
analagously. We will at this point record for later use the following
equation, which follows from \eqgdef. Suppose $x$ and $y$ lie in
$\Sigma_2$ and have the same first term. Suppose also $x\prec y$. Then we
have
$$
m(x,y]=\int_{(\sigma(x),\sigma(y)]}g(x_0z)dm(z).
$$

We will regard the circle as the quotient of the interval [0,1] by the
relation 0=1. Write $o$ for the sequence in $\Sigma_2$ whose terms are all 0.
Now define $\pi:\Sigma_2\ra S^1$ by $\pi(x)=m[o,x]\pmod 1$.
Using elementary properties of $g$-measures (that they are non-atomic and of
full support), we have that $\pi(x)=\pi(y)\Lra x\sim y$.

To check that $\pi$ is surjective, note that $\pi$ is continuous (since
$m$ has no atoms), so that $\pi(\Sigma_2)$ is compact and hence closed.
The set $\pi(\Sigma_2)$ also contains the set $\pi(\{a000\ldots:a$\ \nobreak
is a finite
sequence$\})$, which is dense in $S^1$, so $\pi$ is surjective. We also want
to check that the metric topology on $S^1$ coincides with the quotient
topology it inherits from the projection $\pi:\Sigma_2\ra S^1$. We have
already noted that $\pi$ is continuous with respect to the metric topology
on $S^1$. This implies that the open sets in the metric topology are open in
the quotient topology. We have to check the converse. Suppose $A$ is open in
the quotient topology on $S^1$, that is $\pi^{-1}(A)$ is open in $\Sigma_2$.
This implies that $\pi^{-1}(A)$ is a union of cylinders in $\Sigma_2$.
Pick $\zeta\in A$. Then $\pi^{-1}(\zeta)$ consists of a $\sim$ equivalence
class. If this class has only one member, then since $\pi^{-1}(A)$ consists
of cylinders, it must contain a cylinder which contains $\pi^{-1}(\zeta)$.
It is easy to see that $\zeta$ must be contained in the interior of the
image under $\pi$ of this cylinder, hence $\zeta\in{\rm Int}(A)$.
If the class has two members, then each member must be contained in a
cylinder. These cylinders will project to a left- and a right-neighbourhood
of $\zeta$, which implies, again that $\zeta\in{\rm Int}(A)$. It follows
that $A$ is open in the metric topology, which shows that the two topologies
coincide.

We can use this information to construct the map $T$. Note that if $x\sim y$
then $\sigma(x)\sim\sigma(y)$, so $\pi\circ\sigma(x)=\pi\circ\sigma(y)$.
Using the universal property of quotients, this implies that there is a
continuous map $T:S^1\ra S^1$ such that $T\circ\pi=\pi\circ\sigma$. Now,
$\pi$ is a measure-theoretic isomorphism between the pairs $(\sigma,m)$
and $(T,\pi^*m)$, where $\pi^*m(A)=m(\pi^{-1}(A))$.
Note that $\pi^{-1}(\zeta)$ consists of at most two points. Write
$\rho_+(\zeta)$ for $\max(\pi^{-1}(\zeta))$ and
$\rho_-(\zeta)$ for $\min(\pi^{-1}(\zeta))$. Now, we have
$$
\eqalign{
\pi^*m([0,\zeta])&=m(\pi^{-1}[0,\zeta])
=m([o,\rho_+(\zeta)])\cr
&=\pi(\rho_+(\zeta))=\zeta=\lambda([0,\zeta]).
}
$$
It follows that $\pi^*m=\lambda$, so we have shown that $\pi$ is a
measure-theoretic isomorphism between $(\sigma,m)$ and $(T,\lambda)$.
It remains to show that $T$ is a $C^1$ expanding map. We will in fact show that
$$
T'\bigl(\pi(x)\bigr)=1/g(x).
$$
We will evaluate
$$
\lim_{y\ra x} {{T\bigl(\pi(y)\bigr)-T\bigl(\pi(x)\bigr)}\over
{\pi(y)-\pi(x)}}.
$$
Since $T\circ\pi=\pi\circ\sigma$, this is equal to
$$
\lim_{y\ra x} {{\pi\bigl(\sigma(y)\bigr)-\pi\bigl(\sigma(x)\bigr)}\over
{\pi(y)-\pi(x)}}.
$$
If we now assume $y\succ x$ and that $x$ and $y$
lie in the same 0-cylinder, then the quotient is just
$m(\sigma(x),\sigma(y)]/m(x,y]$. By \eqintg, this converges to
$1/g(x)$ as $y\ra x$ because of the continuity of $g$.
The same analysis can be performed in the case that
$y\prec x$. It is not hard to see that this implies \eqTderiv. Note that the
requirement \eqgclass\ on $g$ is needed to ensure
that the left and right derivatives
coincide at those points $\zeta$ of the form
$\pi(a0111\ldots)=\pi(a1000\ldots)$.

We have that $1/g$ is continuous on $\Sigma_2$ and it collapses equivalence
classes, so we can write $(1/g)=h\circ\pi$ for some continuous function
$h:S^1\ra(1,\infty)$. The above shows that $T'(\zeta)=h(\zeta)$, which
implies that $T\in\E^1$. $\square$
\enddemo

This completes step 1 of the proof, and shows that in order to prove the
theorem, it is necessary only to exhibit a $g$ satisfying \eqgclass, for
which there is a non-ergodic $g$-measure, as the isomorphism described above
then implies that $\lambda$ is non-ergodic with respect to $T$.

We now define a second equivalence relation $\approx$ on $\Sigma_2$.
$$
x\approx y{\rm\ if\ }
\cases{
x=y,&\cr
x=a01000\ldots{\rm\ and\ }y=a11000\ldots&where $a$ is a finite sequence, or\cr
x=a11000\ldots{\rm\ and\ }x=a01000\ldots&where $a$ is a finite sequence.
}$$

\Lemma
Suppose $h\in\G$ has the property that
there is a non-ergodic $h$-measure and
$$
x\approx y\Ra h(x)=h(y)
$$
Then there is a $g\in\G$ satisfying
\eqgclass, such that there is a non-ergodic $g$-measure.
\endproc

\demo{Proof}
Define the 2-1 map $P:\Sigma_2\ra\Sigma_2$ by $P(x)_n=x_n+x_{n+1}$. This map
is certainly continuous. It has two inverse branches $\tau_0$ and $\tau_1$
given by $\tau_0(x)_n=x_0+x_1+\ldots+x_{n-1}\bmod 2$ and
$\tau_1(x)_n=1+x_0+x_1+\ldots+x_{n-1}\bmod 2$. Note that
$\bigl(\tau_i(x)\bigr)_0=i$.
Let ${\cal M}$ denote the probability measures on $\Sigma_2$ and define the
map $\tau^*:{\cal M}\ra{\cal M}$ by
$\tau^*\mu(A)=\sf12\mu({\tau_0}^{-1}A)+\sf12\mu({\tau_1}^{-1}A)$.
This is equal to $\sf 12\mu\bigl(P(A\cap B_0)\bigr) + \sf 12\mu\bigl(P(A\cap
B_1)\bigr)$, where $B_i$ is the 0-cylinder in $\Sigma_2$ of those sequences
whose first term is $i$.
We will use \eqgcons\ to show that if $\mu$ is an $h$-measure,
then $\tau^*\mu$ is an $h\circ P$-measure.
We have for $n\geq 1$,
$$
\eqalign{
{{\tau^*\mu\bigl(\Cyl xn\bigr)}\over
{\tau^*\mu\bigl(\Cyl {\sigma(x)}{n-1}\bigr)}}&=
{{\sf 12\mu\bigl(P(\Cyl xn\cap B_0)\bigr)+
\sf 12\mu\bigl(P(\Cyl xn\cap B_1)\bigr)}
\over
{\sf 12\mu\bigl(P(\Cyl {\sigma(x)}{n-1}\cap B_0)\bigr)+
\sf 12\mu\bigl(P(\Cyl {\sigma(x)}{n-1}\cap B_1)\bigr)}}\cr
&={{\mu\bigl(P(\Cyl xn)\bigr)}\over
{\mu\bigl(P(\Cyl{\sigma(x)}{n-1})\bigr)}}.
}
$$
Since $P$ and $\sigma$ commute and $P(\Cyl xn)=\Cyl{P(x)}{n-1}$,
this is equal to
$\mu\bigl(\Cyl{P(x)}{n-1}\bigr)/$\break$\mu\bigl(\Cyl{\sigma(Px)}{n-2}\bigr)$.
Since $\mu$ is an $h$-measure, we see that
$$
\lim_{n\ra\infty}
{{\tau^*\mu\bigl(\Cyl
xn\bigr)}\over{\tau^*\mu\bigl(\Cyl{\sigma(x)}{n-1}\bigr)}}
=h\circ P(x).
$$
It follows that $\tau^*\mu$ is a $g$-measure, where $g=h\circ P$ as claimed.
However, we see that $x\sim y$ implies that $P(x)\approx P(y)$, so by the
conditions on $h$, we have $g(x)=g(y)$. This means that $g$ satisfies
\eqgclass.
It remains to check that if $\mu$ is non-ergodic, then
$\tau^*\mu$ is non-ergodic. Suppose then that $\mu$ is non-ergodic. There
exists a Borel set $B$ such that $\sigma^{-1}B=B$, with $\mu(B)$ different
from 0 and 1. Since $P$ and $\sigma$ commute, it follows that
$\sigma^{-1}(P^{-1}B)=P^{-1}B$.
Now, we have
$$
\eqalign{
\tau^*\mu(P^{-1}B)&={\textstyle\sf 12}\mu\bigl(P(B_0\cap P^{-1}B)\bigr)+
{\textstyle\sf 12}\mu\bigl(P(B_1\cap P^{-1}B)\bigr)\cr
&={\textstyle\sf 12}\mu(B)+{\textstyle\sf 12}\mu(B)=\mu(B),}
$$
so $\tau^*\mu$ has a shift-invariant set of measure distinct from 0 and 1.
It follows that $\tau^*\mu$ is also non-ergodic as required. $\square$
\enddemo

Note that by combining Lemmas \gtomap\ and \hmeas,
it is now sufficient to exhibit a
function $h\in\G$ satisfying \eqhclass, for which there is a non-ergodic
$h$-measure.

The third step is contained in the following Lemma.

\Lemma
There exists an $h\in\G$ which satisfies \eqhclass\ and for which there
exists a non-ergodic $h$-measure.
\endproc

\demo{Proof}
The proof of this fact is a minor modification of the proof contained in
[BK].
In order to appeal to [BK], however, we have to explain the relationship
between the probabilistic framework used in that paper and the dynamical
framework which we use here.

Bramson and Kalikow consider sequences of random variables $(X_n)_{n\in\Z}$
taking values in the set $\{0,1\}$. They give probabilities for the
evolution of the random variables in the form
$$
\eqalign{
\PP(X_n=1|X_{n-1}=a_{-1},\ X_{n-2}=a_{-2},\ \ldots)&=
\PP(X_0=1|X_{-1}=a_{-1},\ X_{-2}=a_{-2},\ \ldots)\cr
&=f(a_{-1},a_{-2},\ldots).}
$$
They consider stationary distributions for the above process. These are
probabilities on the space of events $\Omega$ for which the conditional
probabilities \eqevo\ hold. The stationarity condition is that
$\PP(X_n=a_0,\ X_{n+1}=a_1,\ \ldots,\  X_{n+k}=a_k)$ is independent of $n$
for a fixed sequence $a$.

We now describe how these probabilities give rise to measures on the space
$S=\prod_{i\in\Z}\{0,1\}$, the space of bi-infinite sequences of 0s and 1s.
Given $\PP$, define a measure $\mu$ on $S$ as follows:
Given a cylinder $\Dcyl amn=\{s\in S:s_i=a_i,\ \forall m\leq i\leq n\}$, let
$\mu(\Dcyl amn)=\PP(X_{-i}=a_i,\ \forall m\leq i\leq n\}$. Note the reversal of
the order. This is necessary because in the probabilistic viewpoint the
terms indexed by negative numbers indicate the past, whereas in the
dynamical viewpoint, the situation is reversed. The stationarity of $\PP$ is
equivalent to the shift-invariance of $\mu$.

There is a natural bijection
between shift-invariant measures on $S$ and shift-invariant measures on
$\Sigma_2$. Given any measure on $S$, it restricts to a measure on
$\Sigma_2$ by composing with the inverse image of the restriction map
$r:S\ra\Sigma_2;\ x\mapsto r(x)$, where $r(x)_i=x_i,\ \forall i\geq 0$.
Conversely, given a shift-invariant measure $\nu$ on $\Sigma_2$,
define $\mu$ on $S$ by $\tilde\nu(\Dcyl amn)=\nu(r(\sigma^{-m}\Dcyl amn))$.
Kolmogorov's extension theorem may then be used to define the measure
$\tilde\nu$ on the whole of $S$. These operations are inverse to one
another. We may therefore regard $\PP$ as giving a shift-invariant
measure on $\Sigma_2$,
the restriction, $m$, of $\mu$ defined above. The condition \eqevo\ on $\PP$
translates to a condition on $m$:
$$
m(\Cyl10|\sigma^{-1}\B)(1a_{-1}a_{-2}\ldots)=f(a_{-1},a_{-2},\ldots){\rm\ for
\ almost\ all\ sequences\ }a{\rm\ (wrt.\ }m).
$$
In other words, $m$ is an $h$-measure, where
$$
\eqalign{
h(1x)&=f(x_1,x_2,\ldots)\cr
h(0x)&=1-f(x_1,x_2,\ldots).}
$$
We now proceed by adapting slightly the proof of [BK].
As in that paper, let $N$ be an integer-valued random variable independent
of $X_{-1},\ X_{-2},\ldots$ taking even values $m_k$ (not odd ones as
previously) with probabilities $p_k$.
Now let $\alpha$ be a sequence $(a_{-1},a_{-2},\ldots)$ as in
[BK]. Using $N$, we define the modified random variable $W$ as follows.
Let $t$ denote those $N-3$ terms of
$(a_{-1},a_{-2},\ldots,a_{-(N-1)})$ which do
not immediately precede the last 1 or the last 0 of $(a_{-1},a_{-2},\ldots,
a_{-N})$. That is $t$ is the word which is obtained by starting with
$(a_{-1},a_{-2},\ldots,a_{N-1})$ and discarding the term which immediately
precedes the last 1 of $(a_{-1},\ldots,a_{-N})$ and similarly discarding the
term preceding the last 0.
Then $W(\alpha)$ is given by
$$
W(\alpha)=\cases{
1-\epsilon&if the majority of the terms of $t$ are 1s\cr
\epsilon&otherwise.}
$$
As in [BK], we define $f(\alpha)={\Bbb E}[W(\alpha)]$, so that $h$ is given
by \eqhdef. The choice of $t$ in the definition of $W$ was made so as to
ensure that $h$ satisfies \eqhclass.
The remainder of the proof in [BK]
\ applies almost verbatim showing that there exists a non-ergodic $h$-measure.
This completes the proof of the Lemma and hence of the Theorem. $\square$
\enddemo

\vskip 0pt plus .3\vsize\penalty -250
\vskip 0pt plus -.3\vsize\bigskip\vskip\parskip
\centerline{R{\ninerm EFERENCES}}
\nobreak\smallskip\noindent

\bigskip
\frenchspacing

\item{[B]}H. Berbee, `Chains with infinite connections: uniqueness and Markov
representation', {\sl Probab. Th. Rel. Fields}, {\bf 76}(1987), 243--253.

\item{[BK]}M. Bramson and S. Kalikow,
`Nonuniqueness in $g$-functions', {\sl Israeli J.
Math.} (to appear).

\item{[GS]}P. Gora and B. Schmitt,
`Un exemple de transformation dilatante et $C^1$ par morceaux de
l'intervalle, sans probabilit\'e absolument continue invariante',
{\sl Ergodic Theory Dynamical Systems}, {\bf 9}(1989), 101--113.

\item{[H]}%
P. Hulse, `Uniqueness and ergodic properties of attractive $g$-measures',
{\sl Ergodic Theory Dynamical Systems}, {\bf 11}(1991), 65--77.

\item{[Ke]}M. Keane, `Strongly mixing $g$-measures', {\sl Invent. Math.},
{\bf 16}(1974), 309--324.

\item{[Kr1]}%
K. Krzy\.zewski, `On expanding mappings', {\sl Bull. Acad. Polon. Sci.
S\'er. Sci. Math.}, {\bf XIX}(1971), 23--24.

\item{[Kr2]}%
K. Krzy\.zewski, `A remark on expanding mappings', {\sl Colloq. Math.},
{\bf XLI}(1979), 291--295.

\item{[KS]}%
K. Krzy\.zewki and W. Szlenk, `On invariant measures for expanding
differentiable mappings', {\sl Studia Math.}, {\bf XXXIII}(1969), 83--92.

\item{[W]}%
P. Walters, `Ruelle's operator theorem and $g$-measures', {\sl Trans. Amer.
Math. Soc.}, {\bf 214}(1975), 375--387.

\end